\documentclass[12pt,preprint]{aastex}
\slugcomment{Submitted to the Astrophysical Journal}
\shorttitle{Spatial Distribution of Satellite Galaxies}
\shortauthors{\'Ag\'ustsson \& Brainerd}

\begin{document}

\title{The Spatial Distribution of Satellite Galaxies Selected from Redshift Space}

\author{Ing\'olfur \'Ag\'ustsson \& Tereasa G.\ Brainerd}
\affil{Boston University, Institute for Astrophysical Research, 725
Commonwealth Ave., Boston, MA 02215}
\email{ingolfur@bu.edu, brainerd@bu.edu}

\begin{abstract}
We  
investigate the spatial distribution of satellite galaxies using a mock redshift
survey of the first Millennium Run simulation.
The satellites were identified using common redshift space criteria and the
sample therefore includes a large percentage of interlopers.
The satellite locations are well-fitted by a combination of a Navarro, Frenk \& White 
(NFW) density profile and a power law.  At fixed stellar mass, the NFW scale
parameter, $r_s$, for the satellite distribution of red hosts exceeds $r_s$ for the satellite
distribution of blue hosts. In both cases the dependence of $r_s$ on host stellar
mass is well-fitted by a power law.  For the satellites of red hosts,
$r_s^{\rm red} \propto (M_\ast / M_\odot)^{0.71 \pm 0.05}$ while for the satellites
of blue hosts, $r_s^{\rm blue} \propto (M_\ast / M_\odot)^{0.48 \pm 0.07}$.  For
hosts with stellar masses $M_\ast \gtrsim 4\times 10^{10} M_\odot$, the
satellite distribution around blue hosts is more concentrated than is the
satellite distribution around red hosts.
The spatial distribution
of the satellites of red hosts traces that of the hosts' halos; however, the 
spatial distribution of the satellites of blue hosts is more
concentrated than that of the hosts' halos by a factor of $\sim 2$.
Our methodology is general and applies to any analysis of
satellites in a mock redshift survey. However, our
conclusions necessarily depend upon the semi-analytic galaxy
formation model that was adopted, and different
galaxy formation models
may yield different results.
\end{abstract}


\section{Introduction}

The spatial distributions of small, faint satellite galaxies have the potential
to place strong constraints on the nature of the
dark matter halos that surround the large, bright ``host'' galaxies about
which the satellites orbit.  
Since the line of sight distances to the vast majority of galaxies are unknown,
studies of satellite galaxies must rely on methods 
other than 
distance to identify the objects of interest.  Traditionally, satellite galaxies
have been identified using proximity criteria that are 
implemented in redshift space, which
necessarily requires both photometry and
spectroscopy. 
The advantages of studies in which the satellites are selected from
redshift space
are: [1] the method identifies a specific set of faint galaxies as
``the'' satellites
and [2] these specific satellites can be linked directly
to the masses of the hosts' dark matter halos via their kinematics.  

When averaged over the entire sample, the satellites of relatively isolated host
galaxies are found preferentially close to the major axes of 
the hosts
(e.g., Sales \&
Lambas 2004, 2009; Brainerd 2005; Azzaro et al.\ 2007; \'Ag\'ustsson \& Brainerd 2010,
hereafter AB10; \'Ag\'ustsson \& Brainerd 2011).
The locations of the satellites
also vary as a function of the physical properties of the hosts
(e.g. Azzaro et al.\ 2007; Siverd et al.\ 2009; AB10). The
locations of the satellites of the host galaxies with the reddest colors and
largest stellar masses are highly-anisotropic, and the satellites
are located preferentially close to the major
axes of their hosts.  The locations of the satellites of the
host galaxies with the
bluest colors and the lowest stellar masses, however, show little to no anisotropy (AB10).
The host
galaxies in these studies typically have stellar masses in the
range $10^{10} M_{\rm sun}$ to $10^{12} M_{\rm sun}$ and, hence, their morphologies are
likely to be ``regular'' (e.g., elliptical, spiral or lenticular).  Therefore,
the trends of observed satellite locations with host physical properties 
are effectively
trends with host morphology.

Using simple assumptions about the ways in which luminous host galaxies
might be embedded within their dark matter halos, \'Ag\'ustsson \& Brainerd (2006b) and
AB10 showed that simulations of 
$\Lambda$-dominated Cold
Dark Matter (CDM) universes can reproduce the observed locations of satellite 
galaxies that were selected from redshift space.  AB10 showed
that the observed trends of satellite locations
with host color and host stellar mass can only be reproduced if
elliptical and non-elliptical (i.e., ``disk'')  hosts are embedded
within their halos in different ways.  In the case of elliptical hosts, AB10 found
that mass and light must be well-aligned,
resulting from a model in which ellipticals
are essentially miniature versions of their dark matter halos.
In the case of disk hosts, AB10 found that mass and
light must be poorly aligned, resulting from a model in which the
angular momentum of the host's disk is aligned with the net angular momentum of
its dark matter halo.  Since the halo angular momentum vectors are not aligned
with any of the halo principle axes (e.g., Bett et al.\ 2007), misalignment
of mass and light in disk hosts therefore occurs.

The kinematics of satellites selected from redshift space 
has also led to important conclusions about the nature of
the dark matter halos surrounding field galaxies, including a broad general 
agreement with the predictions of CDM
(e.g., McKay et al.\ 2002; Brainerd \& Specian 2003;
Prada et al.\ 2003; Conroy et al.\ 2005, 2007; Norberg et al.\ 2008;
Klypin \& Prada 2009; More et al. 2009, 2011). In addition, satellites selected from
redshift space have been shown to be intrinsically aligned with their
hosts (\'Ag\'ustsson \& Brainerd 2006a), an effect which has also been detected 
in numerical simulations (e.g., Velliscig et al.\ 2015).

Here we use a mock 
redshift survey of the
first Millennium Run simulation\footnote{http://www.mpa-garching.mpg.de/millennium}
 (MRS; Springel et al.\ 2005)
to investigate the real space distribution
of satellite galaxies that have been
selected in the same
way one would select satellites from a redshift survey of our 
universe.  The MRS followed the growth of structure in a $\Lambda$CDM cosmology
($H_0 = 73$~km~sec$^{-1}$~Mpc$^{-1}$, $\Omega_{m0} + \Omega_{b0} = 0.25$,
$\Omega_{b0} = 0.04$, $\Omega_{\Lambda 0} = 0.75$, $n=1$, $\sigma_8 = 0.9$) 
from a redshift $z = 127$ to $z=0$ using $N = 2160^3$ dark matter particles of mass
$m_p = 8.6\times 10^8 h^{-1} M_\odot$.  The simulation volume is a cubical
box with periodic boundary conditions and a comoving sidelength of
$L = 500 h^{-1}$~Mpc.  A TreePM method was used to evaluate the gravitational
force law, and a softening length of $5h^{-1}$~kpc was used.

Previous studies of the spatial distributions of satellite galaxies in $\Lambda$CDM
universes have focused on the distribution that results when the satellites are 
selected using 3D information.
Studies of the locations of
satellite galaxies obtained from semi-analytic galaxy formation models
(SAMs)
generally agree that, when selected in 3D, the
satellites trace the dark matter distribution
reasonably well, both within galaxy clusters and around individual, large host
galaxies (e.g., Gao et al.\ 2004; Kang et al.\ 2005; Sales et al.\ 2007).  Using
hydrodynamical simulations, however,
Nagai \& Kravtsov (2005) found that the 
distribution of simulated cluster galaxies had a lower concentration than 
the surrounding cluster dark matter.

Our work differs from that of
Sales et al.\ (2007), who also investigated the spatial distribution of
satellite galaxies in the MRS.  Here
we select our host-satellite systems from redshift space in
the same manner that would be adopted for an observational sample, while
Sales et al.\ (2007)
selected hosts and satellites using full 3D information.
Our sample of hosts and satellites
is completely analogous to an observational sample, including the unavoidable
presence of a significant number of false satellites (a.k.a.\ ``interlopers'') 
in the data.  
Here we aim to address two questions:
[1] How does the spatial distribution of the satellites depend on the
color and stellar mass of the host galaxy?
[2] To what degree does the satellite distribution trace the distribution of
the dark matter surrounding the halos of the host galaxies?
The results presented here draw substantially upon the first author's
PhD dissertation research (\'Ag\'ustsson 2012), and the reader is referred to the
dissertation for more in-depth discussion, as well as complimentary 
figures and tables.

The outline of the paper is as follows.  In \S2, we present the method
we used to select host galaxies and their satellites,
we describe the properties of the 
host-satellite samples that we used in our analysis, and we discuss the way
in which we created stacked ``composite'' host-satellite systems.
In \S3 we compute the spatial distributions of the 
satellites and we determine the best-fitting model parameters for the 
distributions.  In \S3 we also compare the spatial distribution
of the satellites to the spatial distribution of the dark matter surrounding the
host galaxies.
We summarize our results in \S4.

%----------------------------------------------
\section{Host--Satellite Sample}

The satellites in which we are interested are
small, faint 
objects that are located ``close to'' bright objects, both in projected
radial distance
on the sky, $R_p$, and in relative line of sight velocity, $|dv|$.  Here
we use the redshift space selection criteria from AB10, which yield a population of
relatively isolated host galaxies and their satellites. In addition, the
individual host galaxies
dominate the kinematics of the systems.
The selection
criteria require that host galaxies be at least 2.5 times
more luminous than any other galaxy within a projected radial
distance $R_p \le 700$~kpc and a relative line of sight velocity
$|dv| \le 1000$~km~sec$^{-1}$. Satellites must be found within
a maximal projected radial distance $R_{\rm max} =
R_{p,{\rm max}} = 500$~kpc from their host and must
have a
relative line of sight velocity $|dv| \leq 500$~km~sec$^{-1}$.  
Furthermore, we impose a minimum projected radial distance of
$R_{\rm min} = R_{p,{\rm min}} = 25$~kpc for the satellites, since 
few satellites with 
$R_p \le 25$~kpc will be found in an observational
sample (i.e., such nearby satellites are generally
difficult to distinguish from their hosts in the imaging data that are associated with
observational 
redshift surveys.) 
Each satellite must also be
at least 6.25 times fainter than its host.
In order to eliminate a small number of systems that
pass the above tests but which are likely to be
clusters or groups,
we impose two more restrictions: [1]
the luminosity of each host must exceed the sum total of the luminosities
of its satellites, and [2] each host
must have fewer than nine satellites.  The choice of the maximum number of satellites
is somewhat arbitrary and there is no significant effect on our results if we instead choose a
different maximum value of, say, four or five satellites.
This is due to the fact that most of our host galaxies have
only one or two satellites, and the inclusion in the analysis
of a handful of systems with many satellites 
does not affect our results in any significant way.

Our host-satellite sample is taken
from AB10, and we refer the reader to AB10 for additional
discussion of the sample.
The hosts and satellites were obtained by
implementing the above selection criteria using the 
Blaizot2006\_Allsky
mock redshift survey
of the  MRS,
resulting in a total of 65,654 hosts and 130,034 satellites.
The mock redshift survey utilized the De Lucia \& Blaizot (2007) 
SAM and was constructed 
using the Mock Map Facility (MoMaF; Blaizot et al.\ 2005).  The
resulting galaxy catalog is available
on line in Section 3.3.9.2
of the Virgo-Millennium Database.\footnote{
http://gavo.mpa-garching.mpg.de/Millennium/pages/help/HelpSingleHTML.jsp
}
The De Lucia \& Blaizot (2007) SAM is a slightly modified version of the
Croton et al.\ (2006) SAM that was used by Sales et al.\ (2007) 
in their studies of 3D-selected MRS satellites.  The key difference between
the De Lucia \& Blaizot (2007) SAM and the Croton et al.\ (2006) SAM is that
De Lucia \& Blaizot (2007) adopted the Chabrier (2003) initial mass function
and the Padova 1994 evolutionary tracks.

The Blaizot2006\_Allsky
mock redshift survey was designed to reproduce the spectroscopic
selection
of the Sloan Digital Sky Survey (SDSS; e.g., Fukugita et al.\ 1996; Hogg et al.\
2001; Smith et al.\ 2002; Strauss et al.\ 2002; York et al.\ 2000) and, hence, 
it has a relatively
bright limiting AB magnitude of $r=18$.  Because of this bright limiting magnitude,
only one or two satellites are typically identified for
each host galaxy, and the small number of satellites per host
requires that
the spatial distribution of the satellites be investigated using ensemble averages.
We know from AB10 that 94\% of the host galaxies in our sample reside at the centers
of their halos, and therefore it is reasonable to 
stack many individual host-satellite systems together, creating
``composite'' host-satellite systems from which we can determine the average 
spatial distribution of the satellites.
We also note that no specific redshift cuts on the galaxy catalog,
other than the cut that results from the limiting magnitude of $r=18$ (which
was adopted 
for the creation of the mock redshift survey) are used for our sample selection.
Our host galaxies have apparent magnitudes $r \lesssim 16$ 
and median redshifts
$z_{\rm med} \simeq 0.05$ (see Figure~1 of AB10). That is, our sample
is drawn from a set of objects that are brighter and
closer than 
the spectroscopic completeness limits
of both the SDSS and the mock redshift survey.

When creating the composite host-satellite systems, we would ideally stack together 
systems in which the
dark matter halos of the hosts have similar masses. 
In practice, however,
it is not possible to stack
host-satellite systems in an observational sample using
halo mass.
To make the best connection of theory to future observations, we stack
our host-satellite systems using a method that could be implemented straightforwardly
with observational data; since, for simulated galaxies,
stellar mass correlates more strongly with
halo mass than does luminosity
(e.g., More et al.\ 2011 and
references therein) we use the stellar masses of the host
galaxies to stack our systems.
%Shown in the left panel of Figure~1 is the relationship between halo virial
%mass ($M_{200}$) and $r$-band
%absolute magnitude ($M_r$) for our host galaxies.
%Different point types and colors in Figure~1
%indicate different morphologies for the host galaxies, as
%determined from their $B$-band bulge-to-disk ratios (see AB10).  Orange contours indicate
%regions inside which $\sim 95$\% of the red hosts are found and purple contours
%indicate regions inside which $\sim 95$\%
%of the blue hosts are found.  
%From Figure~1, the relationship between halo virial mass
%and stellar mass is much tighter than the relationship between halo
%virial mass and host luminosity, and we therefore use stellar mass to
%stack the host-satellite systems together.
%{\bf We note that Figure~6.1 in \'Ag\'ustsson (2012) conveys the same information
%as Figure~1 of this work, but since the on line version of the
%dissertation is a low resolution, black and white scan of the original document,
%it is difficult to discriminate between the different types of hosts
%in the on line dissertation.  Hence, an improved, full-color
%version of this figure is included here.}

We divide our host sample according to
rest-frame optical
color, $(g-r)$, at redshift $z = 0$.  To do this, we fit the distributions
of the host $(g-r)$ colors by the sum of two Gaussians (e.g., Strateva
et al.\ 2001; Weinmann et al.\ 2006).  
The division between the two Gaussians lies
at $(g-r)=0.75$ and we therefore define ``red'' hosts to be those
with $(g-r) \ge 0.75$ and ``blue'' hosts to be those with
$(g-r) < 0.75$.
Since our host galaxies have a clear bimodal distribution of colors,
we
construct our composite host-satellite systems by stacking
systems in which hosts of a given color are found
within a given stellar mass range.
%Figure~1 shows that
%the hosts with the lowest stellar masses are predominately blue, while the hosts with
%the highest stellar masses are predominately red.  
Here we are primarily
interested in comparing the results for the satellites of
host galaxies with similar stellar masses,
but different colors. Because of this we restrict our analysis 
to host galaxies with stellar masses in the range 
$10.3 \le \log_{10} \left[ M_\ast / M_\odot
\right] \le 11.5$,
for which there are a significant number of both red and blue hosts with similar
stellar masses.
To create the composite host-satellite systems, we adopt a fixed bin width of 0.3~dex in stellar mass
(e.g., comparable to the error in the stellar mass estimates for SDSS galaxies; 
Conroy et al.\ 2009).
Given the stellar mass range for the hosts, 
our analysis concentrates on the satellites of host galaxies 
that are found within one of four stellar mass bins: 
$10.3 \le \log_{10} \left[ M_\ast / M_\odot
\right] <
10.6$ (bin $B_1$), $10.6 \le \log_{10} \left[ M_\ast / M_\odot \right] < 10.9$ (bin $B_2$),
$10.9 \le \log_{10} \left[ M_\ast / M_\odot \right] < 11.2$ (bin $B_3$), and
$11.2 \le \log_{10} \left[ M_\ast / M_\odot \right] \le 11.5$ (bin $B_4$).  When referring
separately to the red or blue hosts within a given stellar mass bin we will use
the notation $B_1^{\rm red}$, $B_1^{\rm blue}$, etc.  Limiting ourselves to these
particular stellar mass bins leaves us with a total of 51,633 hosts and
101,984 satellites for our analysis below.

\begin{deluxetable}{lrrrccc}
\tabletypesize{\scriptsize}
\tablecaption{Red Host Statistics \label{table1}}
\tablewidth{0pt}
\tablehead{
\colhead{  } &
\colhead{  } &
\colhead{  } &
\colhead{  } &
\colhead{Host} &
\colhead{Host} &
\colhead{Host} 
\\
\colhead{Bin} & 
\colhead{$N_{\rm host}$} & 
\colhead{$N_{\rm sat}$} & 
\colhead{$N_{\rm sat}(r_{200})$} &
\colhead{$\log_{10} \left[ M_{\ast,{\rm med}}/M_\odot \right]$} &
\colhead{$\left< r_{200} \right>$ [kpc]} &
\colhead{$\left<  \log_{10} \left[ M_{200}/M_\odot \right] \right>$} 
}
\startdata
%
% the virial radii are in Table 7.1 of Ingi's thesis
%
 B$_{1}^{red}$ &    817 &  1,348 &   551 & 10.53 & 218 & 12.2 \\
 B$_{2}^{red}$ &  5,003 &  9,561 &   4,563 & 10.80 & 272 & 12.5 \\
 B$_{3}^{red}$ & 12,244 & 26,073 &  14,147 & 11.06 & 345 & 12.8 \\
 B$_{4}^{red}$ & 10,863 & 27,220 &  18,087 & 11.32 & 500 & 13.3 \\
\enddata
\end{deluxetable}

\begin{deluxetable}{lrrrccc}
\tabletypesize{\scriptsize}
\tablecaption{Blue Host Statistics \label{table2}}
\tablewidth{0pt}
\tablehead{
\colhead{  } &
\colhead{  } &
\colhead{  } &
\colhead{  } &
\colhead{Host} &
\colhead{Host} &
\colhead{Host} 
\\
\colhead{Bin} & 
\colhead{$N_{\rm host}$} & 
\colhead{$N_{\rm sat}$} & 
\colhead{$N_{\rm sat}(r_{200})$} &
\colhead{$\log_{10} \left[ M_{\ast,{\rm med}}/M_\odot \right]$} &
\colhead{$\left< r_{200} \right>$ [kpc]} &
\colhead{$\left<  \log_{10} \left[ M_{200}/M_\odot \right] \right>$} 
}
\startdata
%
% the virial radii are in Table 7.1 of Ingi's thesis
%
 B$_{1}^{blue}$ &  7,984 &  11,831 & 4,186 & 10.47       & 190 & 12.0 \\
 B$_{2}^{blue}$ &  8,301 &  13,675 & 5,814 & 10.74 & 231 & 12.2  \\
 B$_{3}^{blue}$ &  5,287 &   9,765 & 4,733 & 11.02 & 283 & 12.5  \\
 B$_{4}^{blue}$ &  1,134 &   2,511 & 1,411 & 11.27 & 367 & 12.9  \\
\enddata
\end{deluxetable}

Tables~1 and 2 show 
selected statistics for our host-satellite sample.
Within each stellar mass bin, the columns in Tables~1 and 2 list
the number of hosts, the total number of satellites,
the number of satellites found within the halo
virial radius ($r_{200}$), the median host stellar mass,
the mean halo virial
radius, and the mean halo virial mass
($M_{200}$). 
Despite the similar stellar masses of the red and
blue hosts within a given bin, 
the halos of the red hosts are generally more massive than those
of the blue hosts.
In addition,
each of the stellar mass bins contains hosts
with a wide range of halo virial masses (i.e., the scatter between adjacent bins
is $\sim 0.3$~dex), so there is some overlap between the halo
masses within adjacent stellar mass bins.
Tables~1 and 2 also show that only  52\% of our
satellites are found within the virial radii of
their hosts (58\% of the satellites of red hosts are found within 
the virial radius
 and 42\% of the satellites of blue hosts are found within 
the virial radius). 

Throughout, it should be kept in mind that, although our satellites 
are a subset of all satellites that would be selected using 3D criteria,
the subset is not random.
A sample of satellites 
selected using redshift space criteria
may therefore yield different conclusions than a sample of satellites 
selected using 3D criteria.  
Samples of satellites that are selected
using redshift space criteria are incomplete because of the nature
of the selection criteria.
This incompleteness compared to 3D selection 
is due in part to the 
bright limiting magnitudes of current redshift surveys; i.e.,  only the 
brightest
satellites are selected for any given host galaxy. 
%with the apparent
%magnitude difference between the hosts and satellites being typically in the
%range of $\Delta r \sim 2$ to $\Delta r \sim 2.5$ (see, e.g., \'Ag\'ustsson
%\& Brainerd 2006a). 
Because of this, when Sales et al.\ (2007) used 3D selection criteria to
identify satellites in the MRS,
they found 2.3 times more satellites within their hosts' virial
radii than we find in our sample.
The large, bright
satellites that result from redshift space selection
may have distributions that differ from those of more ``typical'' satellites.
In addition, satellites with large velocities relative to their hosts, but which are
nevertheless bound to their hosts, may be rejected from samples that are selected
in redshift space because the selection criteria impose a
maximum value for the host-satellite relative velocity.  
%It is, of course, possible to increase
%the maximum host-satellite relative velocity
%in order to capture additional, genuine high velocity 
%satellites. However, for our selection criteria, 
%this would come at the expense of significantly increased ``noise'' in the 
%form of a much larger percentage of interlopers in the satellite sample.

%-----------------------------------------------------
\section{Satellite Spatial Distribution}

In this section we investigate
the spatial distribution of our satellites. 
The real space
volume within which our satellites  
are contained is cylindrical. The requirement
that the satellites be found within a relative velocity $|dv| \le 500$~km~sec$^{-1}$
of their hosts results in a distribution 
that is sharply peaked at the locations of the hosts, but extends to large line of sight
distances.  Along the line of sight, the
tails of the satellite distribution extend to $\pm 10$~Mpc relative to the host
galaxies.  The interlopers that reside in the tails of this distribution are 
often treated as a random population 
(see, e.g., McKay et al.\ 2002; Brainerd \& Specian 2003;
Prada et al.\ 2003). However, the vast majority of the interlopers are
members of the local large-scale structure that surrounds the host galaxies. Rather
than being a random population, most interlopers are
intrinsically clustered 
with both the host galaxies and their 
genuine satellites in physical space and in velocity space
(see, e.g., van den Bosch 2004; \'Ag\'ustsson 2012).
We take the origin of
the cylindrical volume to be
the location of the stacked hosts and
we take the axis of the cylinder to be the line of sight
(labelled as coordinate $Z$). 
We take the
projected radial distance, $R_p$, between the hosts and satellites to be the
radial coordinate and the 3D distance to be 
$r = \sqrt{\Delta Z^2 + R_p^2}$, where $\Delta Z$ is the line of sight
distance between the host and satellite.
The probability distributions for the 3D distances between the
hosts and satellites, $P_{\rm 3D}(r)$,
 are shown in Figure~1.  The distributions
peak at small values of $r$
and they extend to radial distances far beyond the
maximum values of $r$ that are plotted in Figure~1.  

\begin{figure}
\begin{centering}
\centerline{\scalebox{0.65}{\includegraphics{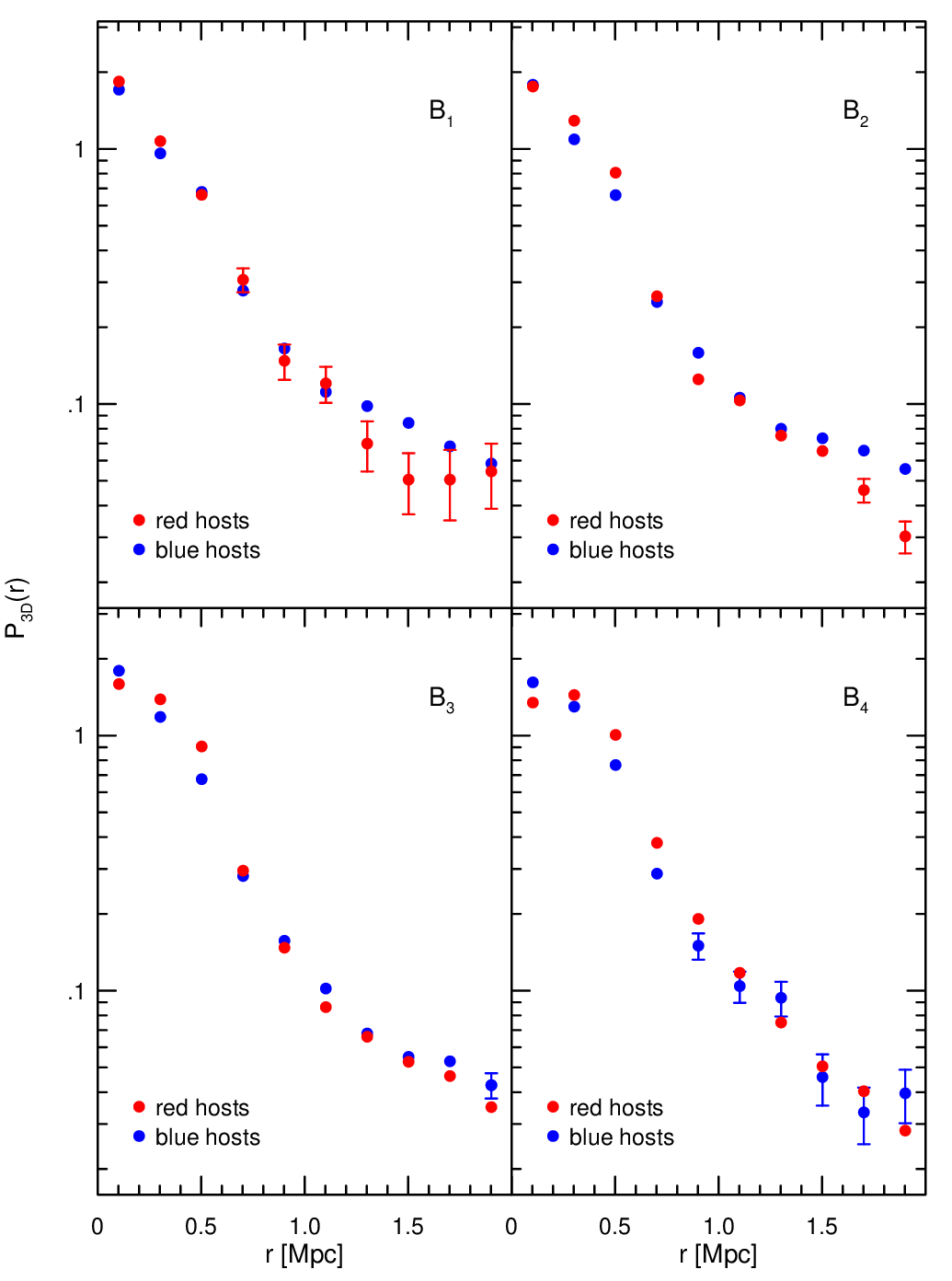}}}
\end{centering}
\vskip -0.4cm
\caption{ 
Probability distributions for the locations of satellites 
surrounding host
galaxies in four different stellar mass bins:
$10.3 \le \log_{10} \left[ M_\ast / M_\odot
\right] <
10.6$ ($B_1$), $10.6 \le \log_{10} \left[ M_\ast / M_\odot \right] < 10.9$ ($B_2$),
$10.9 \le \log_{10} \left[ M_\ast / M_\odot \right] < 11.2$ ($B_3$), and
$11.2 \le \log_{10} \left[ M_\ast / M_\odot \right] \le 11.5$ ($B_4$).
Error bars are standard Poisson errors and are
omitted when they are comparable to or smaller than the 
data points.
Approximately 90\% of the satellites are found within a 3D distance
of $r \le 2$~Mpc from their hosts and $\lesssim 1$\% are found as far as
10~Mpc from their hosts.  Note that the slope of $P_{3D}(r)$ changes at
$r \sim R_{\rm max} = 500$~kpc due to the cylindrical nature of the selection
criteria for satellite distances greater than 500~kpc.
}
\label{fig1}
\end{figure}

%Approximately 90\% of the 
%satellites are found within a 3D distance of $r \le 1$~Mpc from their
%%hosts and
%$\la 1$\% are found as far as 10~Mpc from their hosts.  
%In most cases, the slope of $P_{3D}(r)$ 
%changes at $r \sim
%R_{\rm max} = 500$~kpc, which is due to
%the cylindrical nature of the selection criteria for satellite distances 
%greater than 500~kpc.
%The probability distributions for the 2D satellite locations are shown in 
%Figure~3, from which it is clear that the 2D satellite distribution is
%relatively flat in comparison to the 3D distribution.  This is due
%to the long tails in the 3D satellite distribution that have been integrated
%along the line of sight when producing the 2D distribution. 

Although our satellites are not a random subset of the satellites that would
be obtained if 3D selection criteria were used, we can nevertheless anticipate the 
functional form of the satellite number density profile,
$\nu(r)$, from previous work.
First, we know that our selection criteria yield a sample of
objects that includes both genuine satellites and interlopers.
Sales et al.\ (2007) found that the
number density of 3D-selected MRS satellites (i.e., ``non-interloper'' satellites)
was well-fitted by a Navarro,
Frenk and White profile (NFW; Navarro, Frenk \& White 1995, 1996, 1997) for 
host-satellite distances $r \lesssim 2~ r_{200}$.
Similarly, Kang et al. (2005) found that when satellites of Milky Way-type halos
were selected in 3D, the satellites traced the dark matter distribution well, 
particularly for radii between $0.1 r_{\rm vir}$ and $r_{\rm vir}$.
Therefore
for small values of $r$, we expect $\nu(r) \propto
(r/r_s)^{-1}(1 + r/r_s)^{-2}$, where $r_s$ is the NFW scale radius.
For sufficiently large
values of $r$, the ``satellites'' in our sample are, in fact, interlopers.
On large scales, then,
the form of the probability density for the separation between hosts and satellites should be
the same as the probability density for the separation between galaxies as a whole,
which is approximated well by a power law (e.g., Hayashi \& White 2008).  
Therefore, for large values of $r$ we 
expect $\nu(r) \propto r^{-\gamma}$.

To capture the 3D
distribution of our entire
collection of satellites, we adopt a functional form for the number density
that consists of
a superposition of an untruncated NFW profile and a power law:
\begin{equation}
\nu(r) \equiv \frac{A}{r/r_s \left(1+r/r_s\right)^2} + \frac{C}{r^{\gamma}} \,,
\label{eq:SAT_NFW_WITH_BKG_ORG}
\end{equation}
where $A$ and $C$ are constants. 
On large scales,
the NFW profile falls off as $r^{-3}$ and so long as $\gamma < 3$, the
number density of satellites at large distances 
becomes dominated by the interlopers that reside within the tails of the
distribution.  Throughout, we will refer to these most
distant objects as
``tail interlopers''.  
We also note that 
the NFW profile is technically only valid within
the virial radius, but for convenience
we extend the NFW profile beyond the
virial radii of our hosts in order to smoothly capture the behavior of
$\nu (r)$ at radii that are in between those at which the pure NFW profile and the pure power
law separately dominate.
Below we treat the NFW component of the satellite number counts
separately from that of the tail interlopers.  To do this, we make the definitions
$\nu(r)_{\rm NFW} \equiv A (r/r_s)^{-1} (1+r/r_s)^{-2}$ and
$\nu(r)_{\rm tail} \equiv C r^{- \gamma}$.  Mathematical expressions for the
differential and interior number counts of the satellites in the NFW component
and the tail interlopers are given in the Appendix.

\subsection{Model Parameters for the Satellite Spatial Distribution}

We use a Maximum Likelihood (ML) method to obtain best-fitting model parameters
and the corresponding errors for the satellite number density. We evaluate 
the goodness-of-fit using a combination of the Kolmogorov-Smirnov (KS)
statistic and nonparametric bootstrap resampling (e.g., Wall \& Jenkins 2003;
Babu \& Feigelson 2006).
Throughout, we obtain the bootstrap samples
by repeatedly drawing, with replacements, from the host sample
since the satellites are not all independent of each other (i.e., some
hosts have more than one satellite).  By resampling using the hosts, we insure
that an individual bootstrap sample will contain all host-satellite
pairs for a given host.

In order to implement the ML fitting procedure, we define a probability
density function for the satellite distribution 
of the form
\begin{equation}
P_{\rm 3D}(r) \equiv (1-f_{\rm tail}) P_{\rm NFW}(r) + f_{\rm tail} P_{\rm tail}(r)~,
\end{equation}
where $0 \le f_{\rm tail} \le 1$ is the fraction of tail interlopers.
Here $P_{\rm NFW}(r)$ and $P_{\rm tail}(r)$ are probability density
functions for the NFW component and the tail interlopers, respectively.
%The individual
%probability density functions, $P_{\rm NFW}(r)$ and $P_{\rm tail}(r)$, are
%based on the interpretation of the differential number counts of the
%satellites, $dN$, as the
%unnormalized probability of finding a satellite within the infinitesimal
%interval $\left[ r, r+dr \right]$.
The probability density function has three free parameters:
$r_s$, $\gamma$, and $f_{\rm tail}$.  It also has three geometrical parameters,
two of which are known from the selection criteria:
$R_{\rm min}$ and $R_{\rm max}$.
The third geometrical parameter, $r_{\rm max}$, is the length of the
tail of the satellite distribution in 3D, which is unknown in
an observational sample.
Figure~2 shows the model parameters $r_s$ and $\gamma$
from the ML fits that were computed by adopting
five different values of $r_{\rm max}$ (1, 2, 3.5, 5, and 7~Mpc),
which enclose 83\%, 90\%, 94\%, 96\%, and 98\% of all satellites, respectively.
Errors on the best fit values of $r_s$ and $\gamma$ are the corresponding
standard errors from the maximum likelihood estimates of the parameters, and
are the associated 68\% confidence intervals.

\begin{figure}
\begin{centering}
\centerline{\scalebox{0.8}{\includegraphics{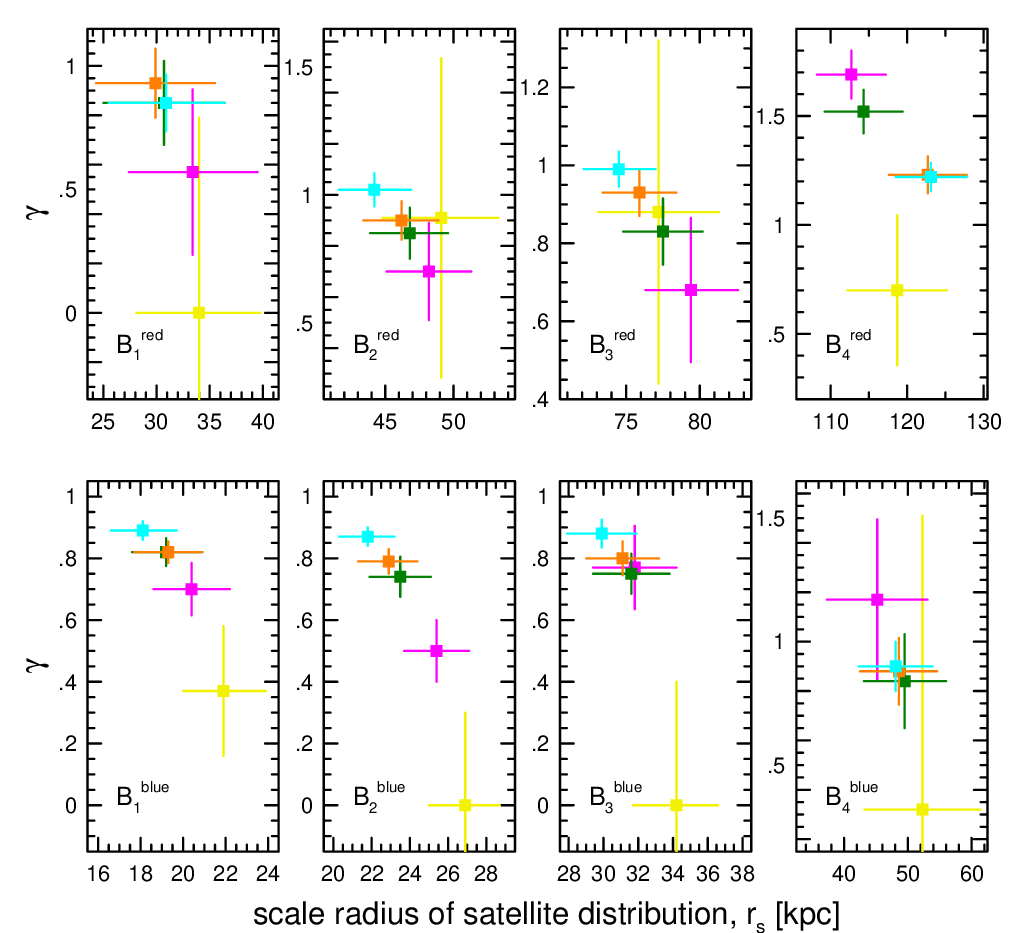}}}
\end{centering}
\caption{Best-fitting model parameters, $r_s$ and $\gamma$, obtained
using different values of $r_{\rm max}$: 1~Mpc (yellow), 2~Mpc (pink),
3.5~Mpc (green), 5~Mpc (orange), and 7~Mpc (blue).  The best-fitting
values of $r_s$ are not particularly sensitive to the value of $r_{\rm max}$
that is adopted, and good model fits require the presence of a tail of 
interlopers in the satellite distribution.  
The best-fitting values of $\gamma$ are weakly-dependent
on the value of $r_{\rm max}$ so long as $r_{\rm max} \ge 3.5$~Mpc (see text).
}
\label{fig2}
\end{figure}

It is clear from Figure~2 that $r_s$ increases with increasing host stellar mass,
the best-fitting value of $r_s$ is not particularly sensitive to the
value of $r_{\rm max}$ that
is adopted and, within a given stellar mass bin, the best-fitting value of $r_s$ is
larger for the satellites of
red hosts than it is for the satellites of blue hosts.
Also, with the exception of the satellites of the
$B_4^{\rm red}$ hosts, the best-fitting value of $\gamma$
is only weakly-dependent on the value of $r_{\rm max}$ so long as $r_{\rm max} \ge 3.5$~Mpc.
In particular, for $r_{\rm max} = 3.5$~Mpc, $\gamma \simeq 0.8$ for all of the satellites
except those
of the $B_4^{\rm red}$ hosts.  
The $B_4^{\rm red}$ hosts have the most massive halos (mean virial mass of $\sim 2.3\times
10^{13} M_\odot$) and, hence, the highest velocity dispersions ($\sigma_v \sim 320$~km~sec$^{-1}$).
Since the selection criteria reject satellites with relative velocities $|dv| > 500$~km~sec$^{-1}$,
it is likely that the sample of satellites around the $B_4^{\rm red}$ hosts is incomplete.
Therefore, in the case of the $B_4^{\rm red}$
hosts, it is likely that our satellite sample is not fully-representative of the 
satellite distribution around these particular hosts.  
We also note that, since $r_s$ and $\gamma$ both originate with model fits to structure
formation in a $\Lambda$CDM universe (for which the mass density field is continuous), $r_s$
and $\gamma$
are not fully independent of one another. Of the two, $r_s$  is the parameter that
we expect to be indicative of the host halo structure, with a strong dependence on
the physical properties of the host galaxies.  
The parameter $\gamma$ primarily reflects the intrinsic clustering of the galaxies.
We expect
$\gamma$ to have at most a weak dependence on the stellar masses of the host
galaxies, and a correspondingly weak correlation with $r_s$.  This is due 
to the fact that the
power law index describing intrinsic galaxy clustering is 
only weakly dependent on the
physical properties of galaxies (see, e.g., Tables~1 and 2
of Zehavi et al.\ 2011).

We quantify the quality of the model fits 
in Figure~2 by computing the
KS statistic, combined with 2,500 bootstrap resamplings, which
provide unbiased estimates of the goodness-of-fit probabilities for
the KS statistics.
From this, the best-fitting models in Figure~2 have KS rejection confidence levels 
$< 76$\%; i.e., in all cases the fits are good and
the goodness-of-fit is unaffected by our choice of the value for
$r_{\rm max}$.  We also note that acceptable fits to the satellite distribution
require the presence of both the NFW and the tail interloper components.
That is, acceptable
fits to the
satellite distribution cannot be obtained if we set $f_{\rm tail} = 0$
(an ``NFW-only'' model) or $f_{\rm tail} = 1$ (a ``power-law only'' model).
In particular, if we attempt to fit the satellite distribution solely by
an NFW component, the resulting model parameter $r_s$
increases monotonically with $r_{\rm max}$ in all of our host
stellar mass bins, rather than converging to the values shown in Figure~2.
Furthermore, the KS statistics show that
the satellite distributions for 
all of our host stellar mass bins are inconsistent with NFW-only models when we include
satellites that lie beyond $r_{\rm max} \gtrsim 2$~Mpc.  Finally, we note
that for the best-fitting NFW $+$ power law models, 
the tail interloper fraction ranges from 
13\% to 24\% for the satellites of the red host galaxies, and 
from 19\% to 30\% for the satellites of the blue host galaxies.

Figure~3 shows the dependence of the best-fitting values of $r_s$ 
on host stellar mass, which is well-fitted by power laws for the satellites 
of both red and blue hosts.
Formally,  $\log_{10} r_s^{\rm red} =
(0.71 \pm 0.05) \log_{10} \left[ M_\ast / M_\odot \right] - (6.0 \pm 0.6)$
for the satellites of the red hosts and
$\log_{10} r_s^{\rm blue}  =
(0.48 \pm 0.07) \log_{10} \left[ M_\ast / M_\odot \right] - (3.8 \pm 0.8)$ for the satellites of
the blue hosts.
From Figure~3, it is also clear that for host stellar masses $M_\ast \gtrsim 4\times 10^{10}
M_\odot$, the satellite distribution around blue hosts is more concentrated
than is the satellite distribution around red hosts, where the concentration is
given by
$c = r_{200}/r_s$. 
It is, of course, important to note that these results are specific to the 
galaxy formation model that was adopted (i.e., De Lucia \& Blaizot 2007)
and, hence, they may not constitute a general result
for galaxy formation in the context of $\Lambda$CDM.

\begin{figure}
\begin{centering}
\centerline{\scalebox{1.0}{\includegraphics{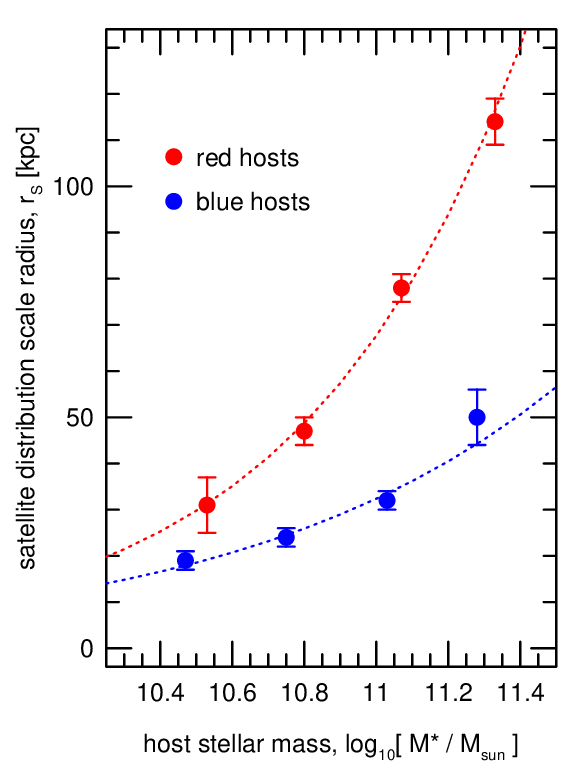}}}
\end{centering}
\caption{Best-fitting values of the scale radius, $r_s$, for the
satellite distributions as
a function of host color and stellar mass.
Here fits to the satellite distributions have been
performed using $r_{\rm max} = 3.5$~Mpc, but we note
from Figure~3 that the value of
$r_s$ is insensitive to the choice of $r_{\rm max}$.  Error bars correspond
to 68\% confidence intervals from the maximum likelihood
parameter estimates. Dotted
lines show the best-fitting power laws for
the dependence of $r_s$ on host stellar mass.  
}
\label{fig3}
\end{figure}

\subsection{Satellite Distribution vs. Dark Matter Distribution}

In addition to simply determining the spatial distribution of the satellites,
we might also hope to use the satellite distribution
to place constraints on the 
concentration of the dark matter halos surrounding the host galaxies.  A direct
constraint on halo concentration would be possible
if the satellites faithfully trace 
the distribution of the dark matter. A number of studies of the
locations of satellite galaxies in SAMs
have led to the conclusion that, when selected in 3D, the satellites
trace the distribution of CDM halos quite well, 
both within galaxy clusters (e.g., Gao et al.\ 2004)
and in lower density environments (e.g., Kang et al.\ 2005; Sales et al. 2007).
Hence, the spatial distribution of 3D-selected satellites would seem to be
a good tracer of dark matter.

We know, however, that our
satellites are not a random subset of all 
luminous satellites.
Since our satellites include only the brightest objects,
it is possible that
the spatial distribution of these particular objects may lead to different conclusions
about the degree to which the satellite distribution traces the distribution of the
underlying dark matter.  For example, Sales et al.\ (2007) investigated the average
satellite luminosity as a function of radius in their 3D-selected sample.
Averaged over their entire sample, Sales et al.\ (2007) found clear
evidence of luminosity segregation, with the average satellite luminosity dropping
by a factor of order 2 between the host centers and their halo virial radii.  Therefore,
faintest and brightest satellites in Sales et al.\ (2007) have different radial distributions on average.

Here we compute the scale radii for the dark matter halos
of our host galaxies and we
compare them to the scale radii of the satellite distributions 
that we obtained above.
To compute the scale radii of the halos,
we use the dark matter particles 
to obtain best-fitting NFW mass profiles.
Within a given host stellar mass bin, the dark matter particles of
the individual hosts were combined into a composite density profile after
appropriately scaling each host by $r_{200}$ and $M_{200}$
(see Chapter~7 of \'Ag\'ustsson (2012).  The
concentration parameters, $c_{dm}$, of the composite hosts were
obtained by fitting NFW profiles, 
and the scale radii were then defined to be the ratios of 
the mean halo virial radii to the concentration parameters:
$r_{s,dm} = \left<r_{200,dm} \right>/c_{dm}$.

\begin{deluxetable}{lcccc}
\tablecaption{NFW Model Parameters for Satellites and Host Dark Matter Halos} 
\tablewidth{0pt}
\tablehead{
\colhead{Bin} & 
\colhead{$r_{s,dm}$} & 
\colhead{$r_{s,sat}$} &
\colhead{$c_{dm}$} &
\colhead{$c_{sat}$} \\
}
\startdata
 B$_{1}^{red}$ & $37.4\pm 0.3$   & $31 \pm 6$   & $5.83\pm 0.05$ & $7.0\pm 1.4$ \\
 B$_{2}^{red}$ & $47.4\pm 0.2$   & $47 \pm 3$   & $5.74\pm 0.02$ & $5.8\pm 0.4$ \\
 B$_{3}^{red}$ & $63.7\pm 0.2$   & $78 \pm 3$   & $5.42\pm 0.02$ & $4.4\pm 0.2$ \\
 B$_{4}^{red}$ & $106.1\pm 0.5$  & $114 \pm 5$  & $4.71\pm 0.02$ & $4.4\pm 0.2$ \\
 & & \\
 B$_{1}^{blue}$ & $36.9\pm 0.1$ & $19 \pm 2$ & $5.15\pm 0.01$ & $10.0\pm 1.1$ \\
 B$_{2}^{blue}$ & $47.1\pm 0.1$ & $24 \pm 2$ & $4.91\pm 0.01$ & $9.6\pm 0.8$ \\
 B$_{3}^{blue}$ & $60.4\pm 0.1$ & $32 \pm 2$ & $4.68\pm 0.01$ & $8.8\pm 0.6$ \\
 B$_{4}^{blue}$ & $79.9\pm 0.2$ & $50 \pm 6$ & $4.59\pm 0.01$ & $7.3\pm 0.9$\\
\enddata
\end{deluxetable}

Figure~4 shows a comparison of
the mean values of $r_s$ obtained for the host halos using the dark 
matter particles 
and the mean values of $r_s$ obtained from the satellites (i.e., 
Figure~3).  
Since the scale radii of the satellite distributions 
increase monotonically with stellar mass, it is clear from 
Figure~4
that the scale radii of the halos also increase monotonically with stellar mass, as 
expected.  From Figure~4, the distribution of satellites around the
red hosts traces the dark matter distribution well.  However, the scale radii for 
the dark matter distributions surrounding the blue hosts exceed those of the 
satellites by a factor of $\sim 2$ in all four stellar mass bins. 
This leads to the conclusion that, when
selected in redshift space, the satellites
of the blue MRS host galaxies have a spatial distribution that is approximately twice
as concentrated as the dark matter surrounding the hosts.
Table~3 summarizes the mean values of the
scale radii of the halo dark matter distributions,
the mean values of the scale radii of the
satellite distributions, and the 
corresponding mean values of the concentration parameters.

\begin{figure}
\begin{centering}
\vskip -0.5cm
\centerline{\scalebox{0.8}{\includegraphics{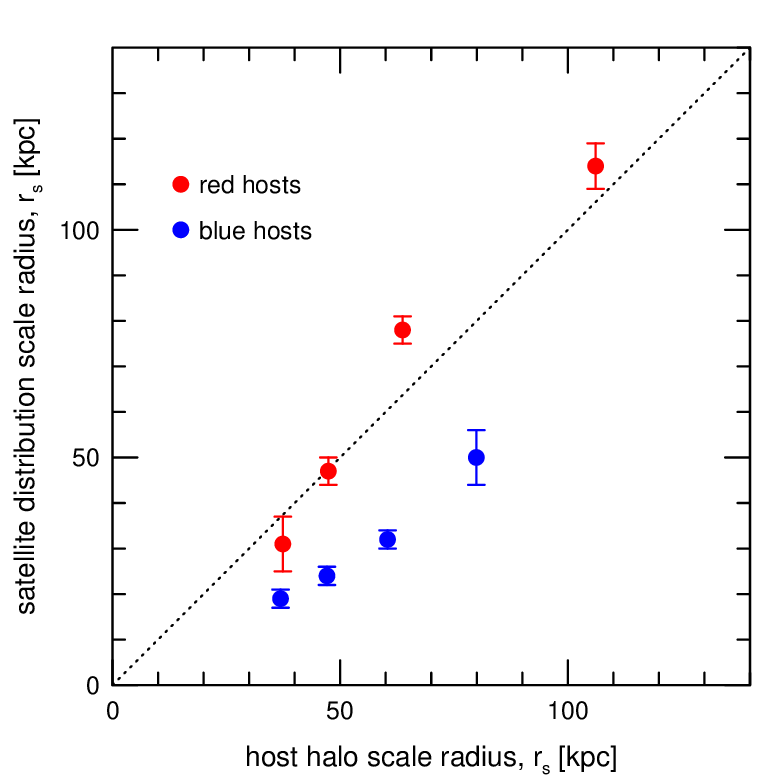}}}
\end{centering}
\caption{Comparison of scale radii for the satellite distributions and
the dark matter halos that surround the hosts.  
The satellites of red hosts are distributed similarly to the dark matter;
the satellites of blue hosts have distributions that are more
concentrated than the dark matter by a factor of $\sim 2$.
}
\label{fig4}
\end{figure}

\begin{figure}
\begin{centering}
\centerline{\scalebox{0.8}{\includegraphics{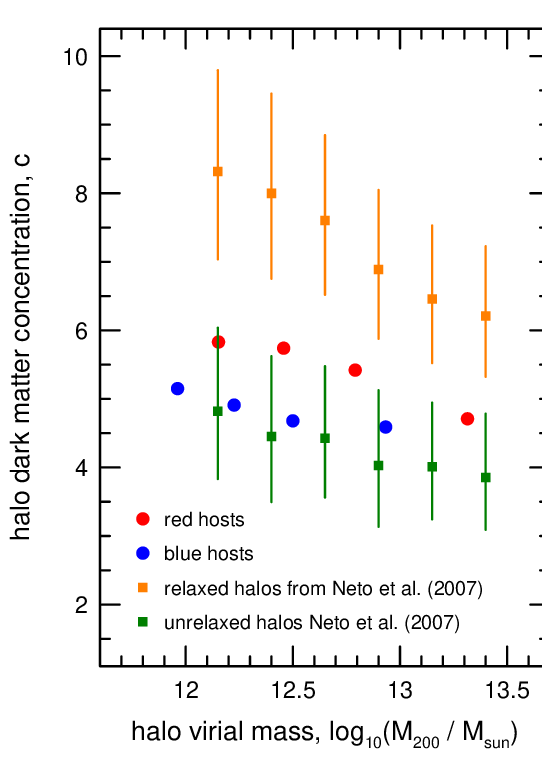}}}
\end{centering}
\caption{Comparison of concentration parameters for the halos of our
host galaxies with those 
obtained for MRS halos by Neto et al.\ (2007).
Error bars for the mean host halo concentration parameters are omitted
because they are smaller than the sizes of the data points.
Vertical lines indicate the ranges of values for
$c$ that correspond to the 25th and 75th
percentiles from Neto et al.\ (2007). 
}
\label{fig5}
\end{figure}

It should be kept in mind that, just as our satellites are not
a random sample of all satellite galaxies, 
our host galaxies are also not a random sample of ``average'' 
galaxies.  
Instead, our hosts are large, bright galaxies that are relatively 
isolated within their local regions of space.  Because of this we would
not necessarily expect the properties of the host dark matter halos to be
identical to the average population in the MRS.  To explore this, we
compare the mean concentration parameters that we obtain for our 
host halos to the mean concentration parameters obtained by 
Neto et al.\ (2007) for all MRS halos.
Shown in 
Figure~5 are the mean values of the concentration parameters
of our host halos as a function of  $M_{\rm 200}$. 
Also shown in
Figure~5 are the mean values of
the concentration parameters and the corresponding 25th to 75th 
percentile ranges for both relaxed and unrelaxed 
 MRS halos from Neto et al.\ (2007).
Within the mass range of the halos
of our host galaxies,
Neto et al.\ (2007) find that a fraction $f_{\rm unrelax} \simeq 0.25$
of all MRS halos are unrelaxed.

From Figure~5, the dependence of the mean 
concentration parameter on $M_{\rm 200}$ for the halos of our
blue hosts is consistent with the results for
unrelaxed MRS halos.  Therefore, it is likely that our sample of blue host
galaxies is dominated by objects with unrelaxed halos, in which case we
would not necessarily expect the spatial distribution of the satellites to trace the
dark matter distribution.  For our red hosts,
the dependence of the mean halo concentration parameter on $M_{\rm 200}$ is
consistent with the upper (75\%) bound on the value of $c$ for unrelaxed MRS halos,
but is slightly below the lower (25\%) bound on the value of $c$ for all MRS halos.  This
suggests that the halos of our red hosts consist of a mixture of relaxed and 
unrelaxed halos, with a larger fraction of the halos of red hosts
being unrelaxed in 
comparison to the complete sample of MRS halos (i.e., $f_{\rm unrelax} > 0.25$
for the halos of our red host galaxies).

To the degree that our results and those of Sales et al.\ (2007)
can be directly compared, we find broad general agreement.
Sales et al.\ (2007) concluded that, when averaged
over their entire 3D-selected sample, 
the distribution of MRS satellites was sightly less concentrated
than the average host dark matter halo ($c_{sat} \sim 5.6$ vs.\
$c_{dm} \sim 8.1$).  They also found that dividing their host
sample in half by either halo mass or host galaxy luminosity resulted in little
change to the inferred concentration for the satellite distribution.
Sales et al.\ (2007) did 
not investigate the concentration of the satellite distribution as a function of
host stellar mass or color, so a direct comparison of our results and theirs is 
limited. 
From Table~3, the concentration of our satellite distributions decreases
somewhat with host halo mass,
%(i.e., although the bins are based on the stellar masses
%of the hosts, the mean host halo mass increases by a factor of $\sim 12.5$ from
%bin $B_1^{red}$ to bin $B_4^{red}$ and by a factor of $\sim 8$ from bin
%$B_1^{blue}$ to bin $B_4^{blue}$; see Tables~1 and 2), 
however the effect is 
marginal (less than $2\sigma$).  Therefore, like Sales et al.\ (2007), we find 
no strong evidence that the concentration of the satellite distribution 
varies as a function of host halo mass.  When we compute the 
mean value of the concentration parameter for our
entire satellite distribution, weighted 
according to the number of satellites within $r_{200}$ in each stellar 
mass bin, the result is similar to the mean value of the concentration
for our entire host dark matter halo sample, weighted by the number
of hosts within each stellar mass bin:  $c_{sat} = 6.3$ vs.\ $c_{dm} = 5.2$.
That is, when averaged over the entire sample, the distribution of
our satellites is slightly more concentrated than is the dark matter
that surrounds the host galaxies; however, the halos of our host galaxies
are somewhat less concentrated than are the halos studied by Sales et al.\ (2007).

%-----------------------------------------------
\section{Summary}

We investigated the spatial locations of the satellites of relatively
isolated host galaxies 
selected from the first
Millennium Run simulation (MRS).  The satellites were identified
using redshift space selection
criteria, in direct analogy with spectroscopically selected satellite galaxies 
in large redshift surveys.  
Our primary conclusions center on the answers
to two questions:
[1] How does the spatial distribution of the satellites depend on the
color and stellar mass of the host galaxy?
[2] To what degree does the satellite distribution trace the distribution of
the dark matter surrounding the halos of the host galaxies?  We summarize the
answers to these questions below.

As expected from previous work, the locations of the 
satellites are well-fitted by a combination
of an NFW profile and a power law.  The NFW profile characterizes the
genuine satellites, while the power law characterizes the interlopers (or
``false'' satellites).
The best-fitting values of the NFW
scale radius, $r_s$, are insensitive to the maximum distance, $r_{\rm max}$, that
we adopt in our analysis.  With the exception of the satellites
of the most massive red hosts,
the best-fitting power law index, $\gamma$,
is also largely insensitive to the value of $r_{\rm max}$ that is adopted, provided
$r_{\rm max} \ge 3.5$~Mpc.  That is, we need a sufficient number of interlopers to be
present in the tail of the distribution in order for the fit to converge to a single value
of $\gamma$.
The best-fitting values of
the NFW scale radius are functions of both the stellar masses and the rest-frame
colors of the host galaxies.  At fixed host stellar mass,
the values of $r_s$ for the satellites
of red hosts exceed those for the satellites of
blue hosts.  The dependence of $r_s$ on host stellar mass is
well-fitted by a power law, but the index of the power law differs for the
satellites of the red and
blue hosts:
$r_s^{\rm red} \propto \left(M_\ast / M_\odot \right)^{0.71 \pm 0.05}$ and
$r_s^{\rm blue} \propto \left(M_\ast / M_\odot \right)^{0.48 \pm 0.07}$.
These relationships constitute predictions for the dependence of the 
spatial distribution of satellite galaxies that should be observed
in our universe, if the satellite galaxies have been selected from
redshift space, if the mass of our universe is dominated
by CDM, and if the adopted 
SAM (i.e., De Lucia \& Blaizot 2007) is 
representative of galaxy formation in our universe.
In particular, we expect the
spatial distribution of the satellites of blue host galaxies 
with $M_\ast \gtrsim 4\times 10^{10} M_\odot$
would be significantly more concentrated than the spatial distribution of the
satellites of
red host galaxies with similar stellar masses.

When we compare the scale radii of the satellite distributions to the 
scale radii of the dark matter halos that surround the host galaxies, we find that
the satellites of red host galaxies trace the dark matter distribution well.
However,
in the case of the satellites of blue host galaxies, the spatial distribution of the
satellites is more concentrated than is the halo dark matter by a factor
of $\sim 2$.  This suggests that, when selected using redshift space
criteria, the spatial distribution of satellite galaxies 
may be a biased tracer of the dark matter distribution that surrounds blue
host galaxies.  From the dependence of the host halo concentration on
virial mass, it is likely that our sample of blue hosts is dominated 
by objects with unrelaxed halos.  Therefore, we would not necessarily expect
the spatial distribution of the satellites of blue host galaxies to 
trace the distribution of the halo dark matter.   We caution, however, that 
this result is based upon the use of a particular galaxy formation model and
therefore may not be a general result for $\Lambda$CDM.  

By their nature, SAMs attempt to model the galaxy formation
process in a manner that is computationally efficient, but which necessarily
relies on simplifying assumptions and a large number of parameters.
Because of the large number of parameters associated with SAMs (i.e., 10 to 30), it is 
challenging to fully explore all of the associated parameter space.  While
highly detailed in their treatment of the relevant physics, 
hydrodynamical simulations of galaxy formation 
are much more computationally expensive than SAMs.  Although a great
deal of progress has been made in recent years, hydrodynamical simulations
must still rely on ``subgrid physics''; i.e., parameterization
on scales smaller than the resolution scale due to the fact that it is not
currently possible to simulate the full range of
physical scales needed to describe galaxy formation
within a large, cosmologically-appropriate volume.  Because
neither method is perfect, neither method provides definitive answers to
all of the questions surrounding the history of galaxy formation. As such, 
SAMs and hydrodynamical simulations are highly complementary tools 
at present.  See, e.g.,
Wechsler \& Tinker (2018) for a comprehensive review.

Because of the complementarity of SAMs and hydrodynamical simulations, and
since no one SAM or hydrodynamical simulation currently provides the 
definitive description of galaxy formation in our universe, 
it will be interesting in the future to compare the
results from this work to similar analyzes of simulations that incorporate
different
SAMs, as well as to the results from high-resolution hydrodynamical simulations.
Other SAMs to which our results could be compared include 
the Guo et al. (2011) SAM for the Bruzual \& Charlot (2003)
and Maraston (2005) population synthesis models.  In particular, Henriques
et al.\ (2012) used the 
Guo et al.\ (2011) SAM 
in their light cones of the
MRS.  The Guo et al.\ (2011) SAM was also used by Guo et al.\ (2013) when
scaling of the growth of structure in the MRS 
to parameters that are consistent with the seven-year Wilkinson
Microwave Anisotropy Probe (WMAP) data. It was also used by
Henriques et al.\ (2015) when updating the Munich galaxy formation model to
the {\it Planck} first-year cosmology.   High-resolution hydrodynamical
simulations to which this work could be compared
include the Illustris
project (e.g., Genel et al.\ 2014; Vogelsberger et al.\ 2014),
the Illustris TNG project
(e.g., Weinberger et al.\ 2017, Pillepich et al.\ 2018), the
GIMIC project (Crain et al.\ 2009), the OWL project (Schaye et al.\ 2010), and 
the EAGLE project
(Schaye et al.\ 2015; Crain et al.\ 2015).
GIMIC makes use of the dark matter structure formation from select regions
of the MRS, while
Illustris, IllustrisTNG, OWLS, and EAGLE are independent of the MRS.

%---------------------------
\section*{Acknowledgments}

We are grateful to the anonymous referee for
constructive comments that significantly improved the manuscript.
We are also grateful for the 
hospitality and financial support of the Max Planck Institute for Astrophysics
that allowed IA to work directly with the MRS particle files in Garching, Germany.
The Millennium Simulation databases
used in this paper, and the web application providing online access to them,
were constructed as part of the activities of the German Astrophysical
Virtual Observatory (GAVO).  This work was partially supported by the National
Science Foundation via grant AST-0708468.

\appendix

\section{Appendix}

Here we present mathematical expressions for the 
differential and interior number counts of the satellites in the NFW component
of the satellite distribution, 
as well as the tail interlopers.
Taking $\nu(r)$ to be the satellite number density profile, the
differential number count is simply the number
of satellites found within the infinitesimal interval $\left[ r, r+dr \right]$,
\begin{equation}
dN(r) =
\int_{S}  \nu(r)~ dS~ dr~
dN(r) = 4\pi \left[ r^2 -r \sqrt{{\cal R}\left(r^2 - R_{\rm max}^2 \right)}\, \right]~
\nu(r)~dr,
\label{eq:DIFF_COUNT_ORG}
\end{equation}
where $S$ is the area of the spherical surface at radius $r$ contained within a
cylinder of radius $R_{\rm max}$  and
${\cal R}(x)$ is the ramp function: ${\cal R} (x) = x$
for $x \geq 0$ and ${\cal R}(x) = 0$ otherwise (see \'Ag\'ustsson 2012).
The interior number count is simply the cumulative number of satellites
that are enclosed within a 3D distance $r$,
\begin{equation}
N(< r) = \int_{0}^{r} dN(r)~.
\label{eq:INTERIOR_COUNT_ORG}
\end{equation}
Using Equations (\ref{eq:DIFF_COUNT_ORG})
and (\ref{eq:INTERIOR_COUNT_ORG}), the 3D differential and
interior number counts of the satellites in the NFW component
are given by
\begin{equation}
dN_{\rm NFW}(r) = 4\pi A r_s ~
\frac{r - \sqrt{{\cal R}(r^2 - R_{\rm max}^2)}}{\left( 1 + r/r_s \right)^2}~dr
\end{equation}
\begin{equation}
N_{\rm NFW}(< r) = 4\pi A r_s^3 \left[
\ln\left( 1 + r/r_s \right) - \frac{r/r_s}{1 + r/r_s} - g_{\rm NFW}(r)
\right]~,
\end{equation}
where
\begin{equation}
g_{\rm NFW}(r) = \left\{ \begin{array}{ll}
0 & \mbox{$r \le R_{\rm max}$} \\
{\rm cosh}^{-1}\left( \frac{r}{R_{\rm max}} \right) -
\frac{\sqrt{r^2 - R_{\rm max}^2}}{r + r_s} -
\frac{r_s~ \cos^{-1}\left(
\frac{r r_s + R_{\rm max}^2}{(r + r_s)R_{\rm max}} \right)}{\sqrt{R_{\rm max}^2 - r_s^2}}
& \mbox{$r > R_{\rm max}$~.}
\end{array}
\right.
\end{equation}
%In the special case of $R_{\rm max} = r_s$, $g_{\rm NFW}(r) = {\rm cosh}^{-1}(r/r_s) -
%2 \left(r^2 - rs^2 \right)^{1/2} \left( r + r_s \right)^{-1}$.
%The total interior number count of the satellites in the NFW component converges and
%is given by
%\begin{equation}
%\lim_{r \rightarrow \infty} N_{\rm NFW}(< r) = 4\pi A r_s^3 \left[
%\frac{r_s}{\sqrt{R_{\rm max}^2 - r_s^2}} \cos^{-1}\left( \frac{r_s}{R_{\rm max}} \right)
%+ \ln\left( \frac{R_{\rm max}}{2r_s} \right)
%\right]~.
%\end{equation}
The 3D differential and interior number counts of the tail interlopers
are given by
\begin{equation}
dN_{\rm tail} = 4\pi C r^{-\gamma}\left[ r^2 - r\sqrt{{\cal R}(r^2 - R_{\rm max}^2)}\,
\right]~dr
\end{equation}
\begin{equation}
N_{\rm tail}(< r) = 4\pi C \left[ (3-\gamma)^{-1} r^{(3-\gamma)} -
R_{\rm max}^{(3-\gamma)} \, g_{\rm tail}(r/R_{\rm max}) \right]~,
\end{equation}
where
\begin{equation}
g_{\rm tail}(x) = \left\{ \begin{array}{lll}
0 & \mbox{$x \le 1$} \\
\frac{1}{2} \left[ x\sqrt{x^2 - 1} - {\rm cosh}^{-1}(x) \right] &
\mbox{$x > 1$, $\gamma = 1$} \\
\frac{1}{(3-\gamma)(\gamma-1)} \left[
\frac{\sqrt{x^2 - 1}}{x^\gamma} \left\{ (\gamma -1)x^2 + 1) \right\}
 + \frac{\gamma}{2} B\left(x^{-2}; \frac{\gamma + 1}{2}, \frac{1}{2} \right)
- \frac{\sqrt{\pi} \Gamma \left( \frac{\gamma + 1}{2} \right)}{\Gamma \left(
\frac{\gamma}{2}\right) }
\right] & \mbox{$x > 1$, $\gamma \ne 1$~.}
\end{array}
\right.
\end{equation}
Here $B(y; a,b) \equiv \int_0^y t^{a-1} \left( 1 - t \right)^{b-1}~ dt$ is the
incomplete Beta function.
These expressions can also be found
in Chapter~7 of \'Ag\'ustsson (2012).
%The total inter number count of the tail interlopers
%is finite for $\gamma > 1$ and is given by
%\begin{equation}
%\lim_{r \rightarrow \infty} N_{\rm tail}(< r) = 2 \pi C (3-\gamma)^{-1} R_{\rm max}^{(3-\gamma)}
%\frac{\sqrt{\pi} \, \Gamma \left( \frac{\gamma-1}{2} \right)}{\Gamma \left(
%\frac{\gamma}{2} \right)}~.
%\end{equation}

\end{document}